\documentclass[11pt,a4paper]{article}

\usepackage[utf8]{inputenc}
\usepackage[T1]{fontenc}
\usepackage[english]{babel}
\usepackage{geometry}
\usepackage{booktabs}
\usepackage{graphicx}
\usepackage{amsmath}
\usepackage{xcolor}
\usepackage[hidelinks]{hyperref}
\usepackage{natbib}
\usepackage{threeparttable}
\usepackage{microtype}
\usepackage{csquotes}
\usepackage{amssymb}

\title{Calibrating the Instrument: Controllability of an LLM-Driven Synthetic Population
}

\author{Mirko Degli Esposti\\
\small Department of Physics and Astronomy, University of Bologna\\
\small \texttt{mirko.degliesposti@unibo.it}}

\date{\today}

\begin{document}
\maketitle

\begin{abstract}
Generative Synthetic Populations (GSP) --- the convergence of
statistically grounded population synthesis, agent-based modelling, and
large language model (LLM) agents --- are attracting growing interest as
tools for urban simulation and institutional communication research.
Before any GSP instrument is used to study a real population, however,
a more basic question must be answered: does it respond to stimuli of
known valence in an ordered, replicable, and group-structured way? We
call this property \emph{controllability}. We ask not whether a
synthetic population tracks humans, but whether it tracks
\emph{itself} --- whether the latent structure we impose on it is
recovered in its own responses. This internal-validity question is
logically prior to any claim about external validity, in the same
sense that characterising an instrument's response function must
precede using it to test a theory.

We report SIVE (Synthetic Instrument Validation Experiment): a
fictional municipality (\emph{Montelago}) and a population of 120
synthetic personas with known latent attitudinal structure are exposed
to seven experimental conditions spanning strongly positive to
strongly negative institutional communications about a municipal water
network. Seven pre-registered criteria, evaluated across a temperature
sweep ($t \in \{0.2, 0.5, 0.7\}$), jointly assess fidelity, stability,
noise floor, specificity, sensitivity, and ordering. All seven criteria
pass at every temperature.

A central finding emerges from a stimulus calibration failure turned
diagnostic success: a message designed to be ``weakly positive'' was
identified by the instrument as functionally negative, traced to
unresolved problems, uncertainty, and institutional passivity in its
text; a redesigned version restored the expected ordering, and its
effect interacts with agents' latent trust in ways not anticipated at
design time. A single-profile noise sub-experiment further shows that
the instrument's intrinsic measurement noise is roughly half the
cross-agent estimate and stable across temperatures --- the instrument
is more precise than a naive cross-agent analysis suggests.

Beyond the aggregate criteria, individual agent trajectories reveal
psychologically coherent micro-dynamics that summary statistics
obscure: agents with near-identical trust deltas under the same
negative stimulus arrive there by qualitatively different routes. We
make the full population and reaction data available via an
interactive explorer
(\url{https://mirko-degli-esposti.github.io/montelago-explorer/}).
\end{abstract}


\section{Introduction: populations as prepared states}
\label{sec:intro}

Cities are among the most instrumented complex systems we have, and among
the least controllable. For a mathematical physicist drawn to urban
questions, the attraction and the frustration are the same: the dynamics
of interest --- how information, trust, and behavioural intentions
propagate through a heterogeneous population embedded in space --- unfold
in systems that admit no true replicates, no controlled perturbations, and
no resets. Agent-based modelling (ABM) has long served as the
computational microscope for such systems, but every microscope needs a
preparation: an initial condition. For social simulation, the initial
condition is a \emph{population}. The question of how to prepare a
population that is faithful to what we know, and maximally noncommittal
about what we do not, is older than the current wave of artificial
intelligence. It is, in Jaynes's sense, a maximum-entropy question
\citep{jaynes1957}. Recent work has shown that
population synthesis under moment constraints from official statistics
can be carried out at scale with the machinery of statistical mechanics
--- Gibbs--Glauber dynamics and persistent contrastive divergence on
Boltzmann-machine-like models \citep{degliesposti2026}; a related
maximum-entropy approach to population synthesis has been developed
independently by \citet{pachet2026}. This paper takes that
standpoint --- populations as prepared states, simulations as
measurements --- and asks what must be validated now that generative AI
has entered the laboratory.

Three research traditions, until recently largely separate, bear on this
question. The first is \emph{synthetic population} (SP) synthesis,
developed over three decades for microsimulation and transport modelling
\citep{harland2012, wu2018,chapuis2022}. Within this tradition, two threads
matter for what follows: classical maximum-entropy and
statistical-mechanics approaches to population synthesis under moment
constraints from official statistics --- as cited above--- and a more recent generative extension via
deep generative models \citep{kim2023} and diffusion-based synthesis
\citep{tang2025}. What all of these populations encode is statistical
structure --- joint distributions of demographic attributes, located in
space. What they deliberately do not encode is an inner life: they
specify who the agents \emph{are}, not how they think. The second
tradition, urban ABM, has supplied the dynamics, but often with
parsimonious cognition: behavioural repertoires too rigid to capture how
real residents interpret, misread, emotionally respond
to, and socially reframe events. The third is the newcomer that appears
to relax precisely this constraint: since the demonstration that LLM
agents equipped with memory, reflection, and planning can produce
believable individual and social behaviour \citep{park2023generative,park2024thousand}, a
fast-growing literature has explored LLM-empowered simulation
\citep{gao2024survey,paglieri2026}. We use the term \emph{generative synthetic
populations} (GSP) for the research programme at this confluence. The
three traditions are complementary in what they lack: classical
synthesis produces populations without minds; ABM produces dynamics with
limited psychology; LLM-based persona generation produces psychology
without statistical control. Each tradition's weakness is another's
core competence.

A word is needed on what we mean, and do not mean, by ``psychology'' in
the previous sentence: a large language model is, mechanically, a
stochastic generator of tokens, with no beliefs, no affect, and no
inner experience. We return to this point, and to the narrative
convention it motivates for describing synthetic personas throughout
this paper, in Section~\ref{sec:micro}.

The long-term vision behind this work is a \emph{generative synthetic
population} in the full urban sense: a population whose statistical
structure is controlled by classical synthesis machinery calibrated on
official statistics, along the lines sketched above, whose individuals
are embedded in a georeferenced urban space with partial knowledge of
their local environment, organised into households and social networks
--- familial, friendship, professional --- and capable of mobility and
peer interaction. Such a population would constitute a genuine urban
digital twin at the individual level: a prepared state of the city from
which counterfactual experiments can be launched, on any question a city
administration might ask --- mobility, housing, public health, or
institutional communication among them.

Within the broad space of urban questions a GSP infrastructure might
eventually address, our own work is currently anchored to one concrete
and narrow case: the impact of institutional communication on public
trust. This is not a topic chosen for convenience. It is motivated by
an ongoing study of institutional risk communication in the
context of the Caffaro contaminated site in Brescia, Italy --- a
national-priority site where decades of PCB contamination have produced
a layered, distrustful public \citep{tarantino2026}. This concrete case is real in
every sense that matters: a real site, a real institutional history, a
real and high-stakes communicative problem, and a synthetic population
grounded in official-statistics marginals for the affected
neighbourhoods. It is exactly the kind of case a mature GSP
infrastructure should eventually be able to inform.

And yet Brescia is precisely the wrong place to ask the question that
must be asked first: is the instrument itself --- independently of any
particular case --- controllable? Does it respond to stimuli of known
valence in an ordered, replicable, and group-structured way? A real,
high-stakes, ethically loaded case is too specific, too overdetermined,
and too noisy to serve as a calibration bench for a question like this.
If a synthetic population behaves oddly under a Caffaro stimulus, we
cannot tell whether this reflects something true about Brescia,
something true about the stimulus we wrote, or simply an artefact of the
underlying generative mechanism --- because we have no independent way
of knowing what the ``correct'' response should look like. Before the
instrument can be trusted on a case where the answer is unknown, it must
be exercised on a case where the answer is known by construction. This
is the physicist's instinct: before you record data with a new detector,
you characterise its response function on sources of known intensity,
in a controlled setting stripped of everything that makes the real
target system interesting and, for calibration purposes, inconvenient.
Only then do you point the detector at the unknown signal.

This is why the present paper does not attempt the full programme
sketched above. It operates instead at the opposite extreme: the
\emph{minimal model}. No real urban context, no spatial embedding, no
mobility, no social network, no inter-agent interaction. One hundred and
twenty synthetic personas respond individually to institutional
messages, in isolation. We adopt this minimal configuration
deliberately, as the direct consequence of the argument just made: a
calibration bench has to be artificial and essential, or it cannot do
its job. We therefore constructed \emph{Montelago}: a fictional
municipality, a synthetic population of 120 personas with a fully known
and withheld latent structure, and a single artificial communication
scenario about a municipal water network. Montelago is essential
precisely because it is not Brescia. It has no real-world referent, no
ambiguity about what the ``true'' answer is, and no ethical weight. What
it offers instead is ground truth: we know, because we designed it, what
each persona's attitude is and how an ordered, well-behaved instrument
ought to respond to stimuli of graded valence. Seven experimental
conditions expose the population to stimuli of known valence; seven
pre-registered criteria measure fidelity, stability, noise floor,
specificity, and sensitivity. This is the only setting in which we can
ask, cleanly, the question posed above --- independently of any question
about Caffaro, or about any real city. This is logically prior to any
claim about external validity  \citep{adornetto2025,larooij2025}, and it is the question this paper
answers.

To the best of our knowledge, the question of controllability ---
whether a GSP instrument responds in an ordered, replicable, and
group-structured way to stimuli of known valence, independently of any
comparison with human behaviour --- has not been posed in these terms
before. The present work is, we believe, a first systematic attempt to
address it. We deliberately bracket here the much broader and harder
question of external validity --- whether a GSP instrument's behaviour
tracks real human behaviour --- which is the subject of a rapidly
growing and largely separate literature. Section~\ref{sec:related}
discusses that literature in some detail and situates the present,
internal-validity contribution relative to it.

We call this experimental system \textbf{SIVE} (Synthetic Instrument
Validation Experiment). We make no claim that it constitutes a
general-purpose protocol transferable to other models or other cases
without modification: the specific design choices described in
Section~\ref{sec:montelago} are tailored to this instrument, this
population, and this stimulus set, and should be understood as an
illustration of an approach rather than a procedure to be applied
verbatim elsewhere. What we do claim is that the underlying question ---
is the instrument controllable before it is trusted? --- is general, and
that the experimental logic used to answer it here can be adapted, case
by case, to other instruments.

Two kinds of results emerge from this exercise. The first is aggregate
and quantitative: seven pre-registered validation criteria (C1--C7) that
together characterise fidelity, stability, noise floor, specificity, and
sensitivity. The second is qualitative and micro-level: individual agent
trajectories that reveal psychologically coherent reactions the
aggregate metrics cannot see. The quantitative results are the core
contribution of this paper and are reported in full. The micro-level
material is, at this stage, described and discussed rather than
systematically analysed; a dedicated treatment of individual-trajectory
analysis is left for separate work, and we are explicit about this
boundary so as not to overpromise what Section~\ref{sec:micro} delivers.

The remainder is organised as follows. Section~\ref{sec:context}
situates SIVE in the context of Caffaro's case and describes the GSP tools we
are using. Section~\ref{sec:related} positions our contribution relative
to the recent literature on LLM agent validation.
Section~\ref{sec:montelago} describes the Montelago experimental system
in detail. Section~\ref{sec:criteria} formalises the validation
criteria. Section~\ref{sec:results} reports the quantitative results.
Section~\ref{sec:micro} presents the micro-level qualitative material.
Section~\ref{sec:discussion} discusses the implications and limitations.
Section~\ref{sec:conclusion} concludes.

\section{Context: Caffaro/Brescia case and the GSP tools}
\label{sec:context}
We describe the Caffaro case in some detail for two reasons. First, it is the
direct applied motivation for SIVE: the instrument characterisation
reported in this paper was prompted by the need to know whether what
we measures is real before interpreting what it finds. Second, it
gives concrete form to the long-term GSP vision sketched in the
introduction: Caffaro/Brescia is precisely the kind of high-stakes,
spatially grounded, socially structured urban problem that a mature GSP
infrastructure should be able to address.
\subsection{The real case}

The Caffaro chemical plant in Brescia operated for decades, releasing polychlorinated biphenyls (PCBs) and other hazardous organochlorine compounds that contaminated the surrounding soils, groundwater, and nearby residential and agricultural areas ( \cite{arpalombardia_caffaro}). The affected communities have experienced a long history of delayed institutional acknowledgement, contested scientific evidence, and postponed remediation. The first visible phase of large-scale remediation works began in February 2026, making institutional communication about the intervention an immediate and consequential challenge (\cite{commissario_caffaro}). The audience is well informed, directly affected by the contamination, and, after decades of delays and unmet expectations, characterized by a high level of institutional distrust.

This combination of high stakes and low
baseline trust is what makes the case an appropriate testbed for
communication experiments.

Using a synthetic population of
residents grounded in official-statistics marginals for the four
neighbourhoods overlapping the contaminated site, we expose alternative
framings of the same municipal communiqué to the same population,
recreated from identical initial conditions for each arm. 
%

Results are reported separately  \citep{tarantino2026}. 
What is relevant here is that the  campaign raised a
methodological question we could not answer within the case itself: how
do we know whether the effects we observe --- or fail to observe --- are
properties of the Caffaro communication problem, or properties of the
instrument? The answer required a separate experiment with a population
of known structure.

\subsection{GSP frameworks in use}

Our primary framework in this case  is AgentSociety
\citep{piao2025}, which organises a simulation around citizen agents
with two-tier memory (structured status memory and episodic stream memory),
modular cognition blocks, and native survey and interview instruments.

 A detailed account of
AgentSociety's strengths and defects as encountered in practice ---
including response-extraction bugs and undocumented pipeline behaviours ---
is available in  \citep{tarantino2026}; here we note only that
those implementation-level findings reinforced the case for a dedicated
instrument characterisation study.

For SIVE, we deliberately moved away from the full AgentSociety agentic
loop. Cross-model testing revealed that the generic tool-decision cycle
of frameworks like AgentSociety inflates LLM call counts by factors of
6--8 for models other than the one used during development (80--100 calls
per agent per condition versus 13 for DeepSeek), with no scientific
benefit for an experiment that has no real environment and no real tools
to use. 
%

\section{Positioning relative to recent work on LLM agent validation}
\label{sec:related}

This section is not a general survey of the GSP landscape --- the three
converging traditions that motivate this work (synthetic population
synthesis, agent-based modelling, LLM-driven persona generation) are
already situated in Section~\ref{sec:intro}. Its purpose is narrower
and more specific: to locate the controllability question this paper
asks relative to a small set of recent contributions that address the
closely related, but logically distinct, question of how LLM-driven
agents relate to human behaviour (\cite{larooij2025}). We contrast our contribution with
four recent positions.

\medskip\noindent\textit{Micro--macro validity gaps.}
\citet{taillandier2025} identify the gap between micro-level
believability --- the capacity of LLM agents to produce locally plausible
behaviour --- and macro-level veridicality as the central unresolved
tension. Their proposed mitigation, Hybrid Constitutional Architectures,
is a structural response to the concern that unconstrained LLM behaviour
may produce plausible-looking but systematically biased macro-outcomes.
SIVE operates at a more primitive level: before asking whether a synthetic
population produces macro-valid outcomes, we ask whether it is
controllable. Internal validity of the kind we test is a necessary
precondition for the calibration work that \citeauthor{taillandier2025}'s
programme requires.

\medskip\noindent\textit{Behavioural coherence of individual agents.}
\citet{mooney2026} use latent-profile analysis to test whether LLM agents
that pass surface-level behavioural screens also satisfy the deeper
consistency requirements of the psychological models they are meant to
instantiate. Their central finding is that surface tests pass while
in-depth coherence tests systematically fail. SIVE operates at a yet more
primitive level: rather than testing coherence against an external
psychological model derived from human data, we test whether a population
with a known, artificially imposed latent structure responds to ordered
stimuli in an ordered and replicable way. The ground truth is the
structure we designed, not a theory derived from human behaviour.

\medskip\noindent\textit{Epistemic status of LLM substitution.}
\citet{hullman2025} provide a formal analysis of when LLM responses can
and cannot substitute for human data. They distinguish heuristic
substitution, statistical calibration via prediction-powered inference,
and exploratory simulate-then-validate strategies. A further finding
directly relevant to our design: LLM response distributions systematically
exhibit lower variance than human distributions, a mechanism that in our
experiments manifests as ceiling and floor effects on trust items under
strongly valenced conditions. SIVE operates outside all three regimes
identified by \citeauthor{hullman2025}: there is no human generating
process to approximate. The formal conditions they identify as necessary
for valid substitution are simply inapplicable when the ground truth is
synthetic and pre-registered. What we term controllability is a logically
prior property.

\medskip\noindent\textit{External fidelity as agent-design problem.}
\citet{liu2026bench} benchmark ten LLMs on twelve canonical human-subject
studies via a Probability Alignment Score. Across all models and agent
designs, alignment ranges from approximately 0.26 to 0.50, largely
independent of model scale and temperature. These results confirm that
external fidelity is structurally limited under current architectures.
They also motivate the question SIVE is designed to answer: if external
fidelity is this unstable, is there at least a more primitive property ---
controllability --- that can be reliably demonstrated?

\medskip\noindent\textit{Summary.}
All four positions share a common thread: the primary concern is how
well LLM-driven agents approximate or align with human behaviour. Our
contribution addresses a more modest, but we argue logically prior,
question. We ask not whether the synthetic population tracks humans,
but whether it tracks \emph{itself} --- whether the structure we impose
on it is recovered in its responses. This is an internal-validity
claim, and it is a necessary first step before any of the external
comparisons above can be meaningfully interpreted. The analogy we find
most useful is from experimental physics: characterising an
instrument's response function is not the same as validating a
physical theory. It is what you do first, before the theory can be
tested at all.

\section{The Montelago experimental system}
\label{sec:montelago}

\subsection{Design rationale}

A key requirement for SIVE was that the latent structure of the population
be known by design and invisible to the model. This rules out any
real-world population, where ground truth is always partial and contested.
We therefore constructed \emph{Montelago}, a fictional Italian municipality
with no real-world referent, and a synthetic population of 120 personas
with explicitly encoded attitudinal structure that is withheld from the
LLM at inference time.

Concretely, each persona is exposed to a short written communiqu\'e ---
modelled on an email or municipal notice that a resident might plausibly
receive --- announcing works or developments related to the local water
network. We refer to each distinct version of this communiqu\'e as a
\emph{condition}: the experimental manipulation consists entirely in
varying the content of this single message across seven versions of
known or hypothesised valence, while holding the channel, the topic, and
the surrounding survey instrument fixed. A persona is exposed to exactly
one condition per administration; differences in response across
conditions are therefore attributable to the content of the message
itself, not to extraneous changes in format or context.


The final design is strictly unidimensional: all conditions concern
the municipal water network (\emph{rete idrica}), and all survey items
measure the same attitudinal constructs. The one exception by design
is the placebo condition (PLA), which is thematically orthogonal to
the water network precisely in order to test specificity;
it is described in Section~\ref{sec:conditions}.

\subsection{Population}

Each persona carries three encoded attitude scores on a 1--10 scale:
institutional trust (\texttt{fiducia\_istituzione}), source credibility
(\texttt{credibilita}), and perceived information adequacy
(\texttt{adeguatezza\_info}). Of these, institutional trust is the
organising variable: the 120 personas are balanced across three latent
groups defined by encoded \texttt{fiducia\_istituzione}:

\begin{itemize}
  \item \textsc{low} ($n=40$): encoded trust $\in [1,3]$ ---
    sceptical or distrustful of municipal institutions
  \item \textsc{med} ($n=40$): encoded trust $\in [4,6]$ ---
    ambivalent, conditionally trusting
  \item \textsc{high} ($n=40$): encoded trust $\in [7,10]$ ---
    consistently trusting, with positive prior experience
\end{itemize}

The other two scores, credibility and information adequacy, are encoded
for each persona alongside trust but do not enter the group definition;
they support the specificity and sensitivity analyses reported in
Section~\ref{sec:results}.

Each persona carries a structured demographic profile (age, gender,
education, occupation, marital status), a one-line dispositional label,
and a background story of approximately 80--120 words written in first
person and anchored to everyday life in Montelago. The background story
is the primary carrier of attitudinal information: it situates the
persona in a specific social position, encodes their history of
interactions with local institutions, and establishes the emotional
register in which they engage with public affairs. Two illustrative
examples:

\begin{quote}
\textit{\textsc{low} group, industrial worker, age 42.} ``I have lived
in Montelago for twenty years. The municipality repaved our street three
years ago and left it half-finished for eight months. When I reported a
gas leak near the school last spring, it took them four days to respond.
I pay my taxes on time and I expect basic services in return. So far,
I have mostly learned not to expect too much.''
\end{quote}

\begin{quote}
\textit{\textsc{high} group, retired civil servant, age 67.} ``I worked
for the municipality for thirty years and I know how hard the staff try,
often with very little. When my wife fell ill, the social services
department supported us beyond what was required. Montelago is not
perfect, but it is administered by people who care. I follow the council
meetings and I trust that decisions, even when I disagree, are made in
good faith.''
\end{quote}

\begin{figure}[h]
\centering
\includegraphics[width=0.72\textwidth]{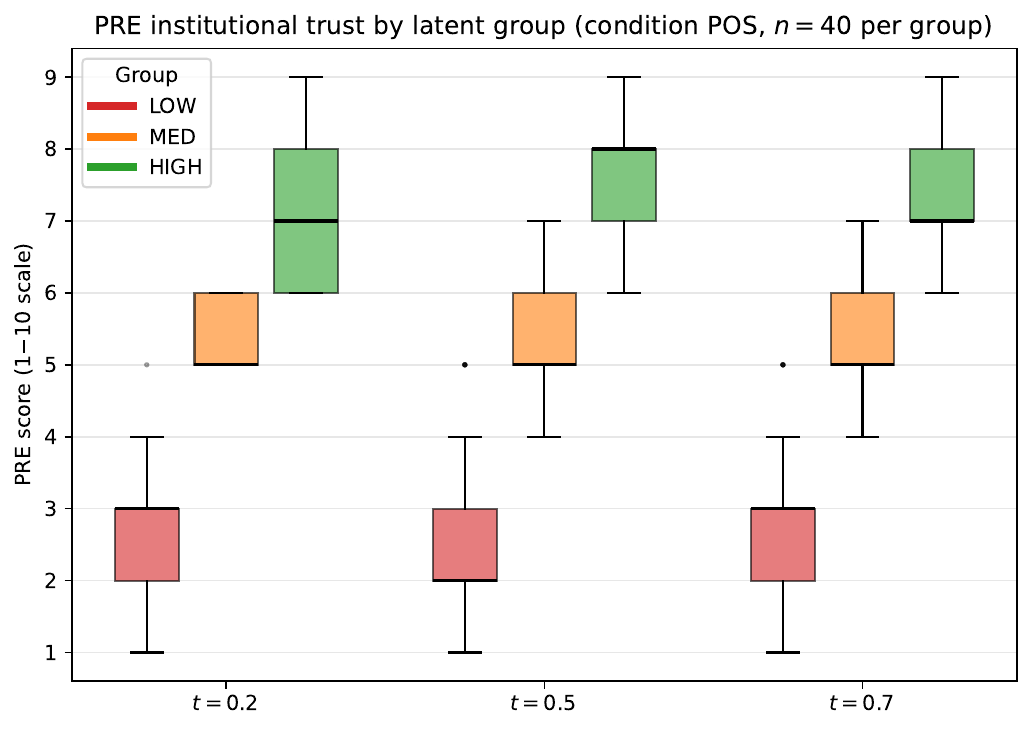}
\caption{Distribution of PRE \texttt{fiducia\_istituzione} scores by
latent group (LOW / MED / HIGH) across the temperature sweep,
condition POS, $n=40$ per group. Boxes show interquartile range;
horizontal lines show medians. The three groups are well separated
and stable across temperatures, confirming monotone recovery of the
latent structure (C1).}
\label{fig:population_pre}
\end{figure}

Figure~\ref{fig:population_pre} shows the distribution of PRE
\texttt{fiducia\_istituzione} scores by group across the temperature
sweep, confirming that the three groups are well separated and stable.

Crucially, all three encoded scores appear only in the metadata layer
accessible to the analyst; the LLM receives only the demographic profile
and background story. The fidelity criterion C1 is therefore a genuine
end-to-end test of the profile $\to$ memory $\to$ response pipeline, not
a recall task: the model cannot ``read off'' the encoded values because
they are never shown to it.

The full population of 120 personas, with complete background stories
and reaction transcripts for all seven experimental conditions, is
available at \url{https://github.com/mirko-degli-esposti/montelago-explorer}
and browsable interactively at
\url{https://mirko-degli-esposti.github.io/montelago-explorer/}. 

\subsection{Experimental conditions}

\label{sec:conditions}

Each persona is administered seven conditions in independent runs
(within-subjects, full reset between conditions), instantiating the
general notion of \emph{condition} introduced in
Section~\ref{sec:montelago}. The design is deliberate: POS and NEG
define the poles of the valence space; POS2 provides an exact replica
of POS to measure the noise floor; POSW and POSW2 probe sensitivity to
weak signals and the instrument's capacity to detect stimulus
miscalibration; PLA tests specificity via a thematically orthogonal
placebo; CTRL separates message effects from re-administration effects.
Together, the seven conditions support all seven validation criteria.
\begin{description}
  \item[POS] Positive pole, defines one extreme of the valence space.
  \item[POS2] Exact replica of POS, for noise-floor estimation (C3).
  \item[POSW] Weak positive signal; \emph{post hoc} found to behave as
    functionally negative (discussed below and in
    Section~\ref{sec:results}).
  \item[POSW2] Recalibrated weak positive, addressing the defects
    identified in POSW.
  \item[NEG] Negative pole, defines the other extreme of the valence
    space.
  \item[PLA] Thematically orthogonal placebo, for specificity (C4).
  \item[CTRL] No message, isolates re-administration effects.
\end{description}
The stimuli are written in Italian, the language of the personas and
their background stories. All five message conditions are reproduced
below in their original form, with English translations provided in
footnotes.\footnote{Translations are functional rather than literal,
preserving register and pragmatic force.}

\begin{quote}
\textbf{MSG\_POS.} \textit{``Il Comune di Montelago comunica che i
lavori di sostituzione completa della rete idrica nel quartiere
inizieranno luned\`i prossimo. Il progetto, finanziato con fondi
europei, prevede la posa di nuove tubature su tutti i tratti principali
e si concluder\`a entro 60 giorni. La qualit\`a dell'acqua sar\`a
monitorata quotidianamente da tecnici certificati e i risultati saranno
pubblicati ogni settimana sul sito del Comune. Eventuali interruzioni
del servizio saranno comunicate con almeno 48 ore di anticipo. Il
Comune garantisce un numero verde attivo h24 per segnalazioni e
informazioni.''}\footnote{``The Municipality of Montelago announces
that works to completely replace the water network in the neighbourhood
will begin next Monday. The project, funded by European grants,
involves laying new pipes along all main sections and will be completed
within 60 days. Water quality will be monitored daily by certified
technicians and results will be published weekly on the municipal
website. Any service interruptions will be notified at least 48 hours
in advance. The Municipality guarantees a freephone line active 24
hours a day for reports and information.''}
\end{quote}

\begin{quote}
\textbf{MSG\_NEG.} \textit{``Il Comune di Montelago comunica che, a
causa di ripetuti guasti alle tubature della rete idrica, si
verificheranno interruzioni del servizio idrico nel quartiere per
almeno i prossimi tre mesi. I lavori di riparazione non possono essere
avviati prima dell'autunno per mancanza di fondi. Nelle ultime analisi,
i valori di torbidit\`a dell'acqua hanno superato i limiti consigliati
in alcune vie della zona. Il Comune si scusa per i disagi.''}\footnote{
``The Municipality of Montelago announces that, due to repeated failures
of the water network's pipes, water service interruptions will occur in
the neighbourhood for at least the next three months. Repair works
cannot begin before autumn due to lack of funds. In the latest analyses,
water turbidity values have exceeded recommended limits on some streets
in the area. The Municipality apologises for the inconvenience.''}
\end{quote}

The contrast between POS and NEG is maximal by design: every dimension
of the communication --- action taken, timeline, transparency,
competence, and institutional affect --- points in opposite directions.
This ensures that C5 (sensitivity) and C6 (ordering) have a strong
signal to detect.

\begin{quote}
\textbf{MSG\_PLA.} \textit{``Il Comune di Montelago comunica che sabato
prossimo si terr\`a in Piazza Centrale la Festa del Quartiere, con
mercatino dell'artigianato locale, musica dal vivo e attivit\`a per
bambini. L'evento \`e organizzato in collaborazione con le associazioni
di volontariato locali e avr\`a inizio alle ore 10:00. L'ingresso \`e
libero e gratuito per tutti i cittadini.''}\footnote{``The Municipality
of Montelago announces that next Saturday the Neighbourhood Festival
will take place in Piazza Centrale, with a local artisan market, live
music, and activities for children. The event is organised in
collaboration with local volunteer associations and will begin at
10:00. Admission is free and open to all residents.''}
\end{quote}

PLA is constructed to be thematically orthogonal rather than merely
``positive'': it makes no reference to the water network, to municipal
competence, or to institutional trust in any form. If the instrument is
specific --- if it responds to the \emph{content} of a stimulus rather
than merely to its presence --- exposure to PLA should leave trust-related
measures statistically indistinguishable from CTRL (criterion C4).

The two weak-signal conditions are the most delicate part of the design,
and one of the paper's main findings concerns what their comparison
reveals about the instrument.

\begin{quote}
\textbf{MSG\_POSW.} \textit{``Il Comune di Montelago informa che sono
in corso valutazioni tecniche per un possibile intervento di
miglioramento della rete idrica nel quartiere. Non sono ancora
disponibili tempistiche definite n\'e dettagli sulle modalit\`a di
intervento. L'amministrazione comunale seguir\`a l'evoluzione della
situazione e comunicher\`a eventuali aggiornamenti.''}\footnote{``The
Municipality of Montelago informs residents that technical evaluations
are under way for a possible improvement to the water network in the
neighbourhood. No definite timeline or details on the form of the
intervention are yet available. The municipal administration will
follow developments and will communicate any updates.''}
\end{quote}

POSW was designed, and initially labelled, as ``weakly positive'': it
announces no problem and gestures at a possible future improvement.
\emph{Post hoc} analysis (Section~\ref{sec:results}) showed that the
instrument treats it as functionally negative. On inspection, the
reason is not hard to see, and it is informative rather than
embarrassing: POSW commits to no concrete action (``valutazioni
tecniche'', evaluations, not works), offers no timeline (``non sono
ancora disponibili''), and closes on a posture of institutional
passivity (``seguir\`a l'evoluzione della situazione'') rather than
commitment. For a persona already inclined to scepticism, a message
that promises nothing concrete is not neutral --- it reads as
confirmation of exactly the pattern (vague reassurance, no
accountability) that scepticism is calibrated to detect. This is a
finding about what counts as positive valence for a distrustful
audience, not an instrument failure; we return to its implications for
applied risk communication in Section~\ref{sec:discussion}.

\begin{quote}
\textbf{MSG\_POSW2.} \textit{``Il Comune di Montelago comunica che \`e
stato avviato un monitoraggio periodico della rete idrica nel
quartiere, come parte delle attivit\`a ordinarie di manutenzione
preventiva. L'amministrazione verifica regolarmente il funzionamento
dell'impianto e resta disponibile a fornire informazioni ai cittadini
che ne facciano richiesta.''}\footnote{``The Municipality of Montelago
announces that periodic monitoring of the water network in the
neighbourhood has been initiated, as part of routine preventive
maintenance activities. The administration regularly checks the
functioning of the system and remains available to provide information
to residents who request it.''}
\end{quote}

POSW2 was constructed to repair exactly the two defects identified in
POSW, while remaining a deliberately low-intensity, low-affect
stimulus: it reports a concrete, already-initiated action (``\`e stato
avviato'', has been initiated, rather than ``sono in corso
valutazioni'', evaluations are under way), and it replaces the passive
closing posture with a standing commitment to availability (``resta
disponibile a fornire informazioni''). No promises, no timeline, no
appeal to affect --- but also no admission of uncertainty or inaction.
As reported in Section~\ref{sec:results}, this recalibration restores
the expected positive ordering, confirming that the original POSW
finding was a property of the stimulus, not of the instrument.

\subsection{Measurement instrument}

Each condition administration follows a fixed three-phase sequence.
In the \textsc{pre} phase, the agent is asked five questions about
their current attitudes toward the municipality and its communication
--- before any stimulus has been delivered. The agent then receives
the condition stimulus (or silence, for \textsc{ctrl}) and produces
three free-text reactions at fixed temporal markers: an immediate
reaction, a reflection after ``a couple of hours have passed'', and a
final reflection after ``the rest of the day has passed''. These
reactions are appended sequentially to the narrative context, allowing
the agent's interpretation of the message to develop over a simulated
day. Finally, in the \textsc{post} phase, the same five questions are
repeated. The agent's post-stimulus ratings therefore reflect not the
message alone but its processing over time. The difference
POST$-$PRE is the primary outcome measure throughout this paper; the
reaction transcripts are themselves a primary output of the experiment
and are the subject of the qualitative analysis in
Section~\ref{sec:micro}.

The five survey items are identical in PRE and POST. The first three
are numeric ratings on 1--10 integer scales:
\begin{enumerate}
  \item \textit{Institutional trust} (\texttt{fiducia\_istituzione}):
    ``How much do you trust the municipality of Montelago in managing
    the water network?'' (1 = no trust, 10 = full trust)
  \item \textit{Source credibility} (\texttt{credibilita}):
    ``How credible do you find the municipality's communication about
    the water network?'' (1 = not at all, 10 = completely)
  \item \textit{Information adequacy} (\texttt{adeguatezza\_info}):
    ``How adequate do you find the information you have received about
    the water network?'' (1 = completely inadequate, 10 = very adequate)
\end{enumerate}
The remaining two items are categorical:
\begin{enumerate}
  \setcounter{enumi}{3}
  \item \textit{Emotion} (\texttt{emozione}): single choice among five
    options --- relief (\textit{sollievo}), worry
    (\textit{preoccupazione}), anger (\textit{rabbia}), hope
    (\textit{speranza}), indifference (\textit{indifferenza}).
  \item \textit{Behavioural intention} (\texttt{intenzione}): single
    choice among five options --- seek more information, talk to
    neighbours, attend public meetings, nothing will change for me,
    contact the municipality.
\end{enumerate}

The three numeric items map directly onto the three encoded attitude
scores in the persona metadata, making them the primary vehicle for
criteria C1 (fidelity), C3 (noise floor), C4 (specificity), C5
(sensitivity), and C6 (ordering). The categorical items provide
qualitative corroboration and are analysed separately for
distributional patterns and entropy (Section~\ref{sec:results}).

Items are administered as \emph{side queries}: each question is
appended to the agent's current narrative context, the response is
recorded, but is not inserted back into the narrative stream. This
design choice has two consequences. First, the battery does not
contaminate the unfolding reaction sequence --- the agent's emotional
and cognitive state evolves freely between \textsc{pre} and
\textsc{post} without being anchored by its own earlier numeric
responses. Second, the call count is fixed at exactly 13 per agent per
condition (5 \textsc{pre} + 3 reactions + 5 \textsc{post}),
regardless of model or temperature, making cross-condition comparisons
exact.

\subsection{Technical implementation}

The SIVE harness runs on Google Colab with incremental Drive
checkpointing after each condition. The primary LLM backend is DeepSeek
(\texttt{deepseek/deepseek-chat}) via OpenRouter, accessed through a
direct \texttt{openai.AsyncOpenAI} client. All results are reported
across a temperature sweep $t \in \{0.2, 0.5, 0.7\}$; we adopt
$t=0.2$ as the reference configuration throughout the paper's narrative
and as the sole temperature used in the interactive online demo
(\url{https://mirko-degli-esposti.github.io/montelago-explorer/}).
A robustness analysis of the sweep is discussed in
Section~\ref{sec:discussion}.
\section{Validation criteria}
\label{sec:criteria}

We specify seven criteria in advance, fixed before inspection of the
results, organised around the analogy of instrument characterisation.
Each criterion is evaluated across the full temperature sweep
$t \in \{0.2, 0.5, 0.7\}$ (Section~\ref{sec:results}); thresholds below
are stated once and apply uniformly across temperatures. The primary
outcome variable throughout is \texttt{fiducia\_istituzione}; secondary
analyses extend to \texttt{credibilita} and \texttt{adeguatezza\_info}
wherever reported. The one exception is C7, where
\texttt{adeguatezza\_info} --- perceived information adequacy --- is
itself the primary outcome, since C7 specifically tests whether agents
register having received information at all, rather than testing the
direction or magnitude of an attitude shift.

\paragraph{C1 --- Persona fidelity.}
The Pearson correlation between encoded attitudinal values
(\texttt{fiducia\_istituzione} in the metadata) and PRE survey
responses, computed across all 120 agents. PRE responses are taken
from condition POS by convention; C2 (cross-replica stability)
confirms that PRE measurements are effectively interchangeable across
conditions, so the choice is immaterial.
\emph{Threshold}: $r > 0.80$. This tests whether the profile $\to$
memory $\to$ response pipeline transmits the latent structure
faithfully.

\paragraph{C2 --- Cross-replica stability.}
For each condition, collect the PRE responses on
\texttt{fiducia\_istituzione} across all 120 agents into a single
vector. Because PRE is measured before any stimulus is delivered, an
agent's PRE score should not depend on which condition it happens to
precede: the seven per-condition vectors are, in principle, seven noisy
measurements of the same underlying quantity. C2 is the minimum Pearson
correlation across all $\binom{7}{2}=21$ pairs of these vectors:
low cross-condition correlation would indicate that PRE measurements
drift depending on experimental context, undermining the within-subjects
design. \emph{Threshold}: $r_{\min} > 0.85$.

\paragraph{C3 --- Noise floor.}
POS and POS2 are two independent administrations of the identical
stimulus \textsc{MSG\_POS} to the same 120 agents, run as separate
sessions with a full reset in between (Section~\ref{sec:montelago}).
Each administration has its own PRE and POST measurement, so each
agent yields two independent deltas on \texttt{fiducia\_istituzione},
$\Delta_{\text{POS}} = \text{POST}_{\text{POS}} - \text{PRE}_{\text{POS}}$
and $\Delta_{\text{POS2}} = \text{POST}_{\text{POS2}} -
\text{PRE}_{\text{POS2}}$. We work with deltas rather than raw POST
scores to control for each agent's individual baseline. Because the
stimulus is identical in both administrations, any systematic
difference between an agent's two deltas cannot reflect a real
treatment effect: it can only be a symptom of intrinsic variability in
the generative process --- the instrument's noise. C3 quantifies this
noise through the per-agent distribution of
$(\Delta_{\text{POS}} - \Delta_{\text{POS2}})$, and is reported in two
complementary parts.

\textbf{C3a} (noise floor): the standard deviation
$\sigma_{\text{noise}}$ of this distribution, reported as a quantitative
instrument characteristic analogous to measurement resolution --- not a
binary pass/fail, but a number that enters the signal-to-noise ratio
used elsewhere in the paper. This cross-agent estimate convolves true
instrument noise with genuine interpersonal variability in how
different agents respond to repeated administration; a single-profile
repetition sub-experiment (Section~\ref{sec:results}), in which a fixed
agent is queried many times, isolates the instrument-noise component
alone.

\textbf{C3b} (replica bias): the mean of
$(\Delta_{\text{POS}} - \Delta_{\text{POS2}})$ across agents should be
practically negligible --- if it were not, repeated administration of
an identical stimulus would itself be introducing a systematic shift,
which would undermine the within-subjects logic of the entire design.
We adopt a region of practical equivalence (ROPE) of
$|\text{mean}| < 0.5$ scale points (5\% of the 1--10 range) rather than
a null-hypothesis significance test, because at $n=120$ a $t$-test has
sufficient power to reject mean~$=0$ for scientifically irrelevant
biases. \emph{Threshold}: $|\text{mean diff}| < 0.5$.

\paragraph{C4 --- Specificity (placebo null effect).}
The mean of $({\Delta}_{\text{PLA}} - {\Delta}_{\text{CTRL}})$ should be
practically negligible. As in C3b, we adopt a ROPE criterion rather than
a significance test, for the same reason: at $n=120$, a $t$-test has
ample power to reject mean~$=0$ for a bias too small to matter. We use
the same region of practical equivalence, $|\text{mean diff}| < 0.5$
scale points, applied here to $\texttt{fiducia\_istituzione}$.
\emph{Threshold}: $|\text{mean diff}| < 0.5$. This tests whether the
placebo is genuinely inert.

\paragraph{C5 --- Sensitivity.}
For each condition, compute the signed delta
$\Delta = \text{POST} - \text{PRE}$ on \texttt{fiducia\_istituzione},
averaged across the 120 agents. C5 asks whether these condition-level
means are ordered consistently with the known valence of the stimuli.

The design was originally pre-registered with six conditions, before
POSW2 existed, with expected ordering
$$\text{POS} \approx \text{POS2} > \text{POSW} > \text{CTRL} \approx
\text{PLA} > \text{NEG},$$
POSW being intended as a weak positive signal, expected to fall between
the strongly positive pole and the CTRL/PLA baseline. As detailed in
Section~\ref{sec:results}, \emph{post hoc} analysis of the POSW
stimulus text revealed it to behave as functionally negative rather
than weakly positive. In response, we constructed a recalibrated
stimulus, POSW2, as a seventh condition, designed to repair the
defects identified in POSW while preserving its intended weak-positive
character. The ordering actually evaluated against the data, and
reported throughout the paper, is therefore the revised seven-condition
ordering
$$\text{POS} \approx \text{POS2} > \text{POSW2} > \text{CTRL} \approx
\text{PLA} > \text{POSW} > \text{NEG}.$$
We report both orderings for transparency: the discrepancy between the
original six-condition expectation and the revised seven-condition one
is itself one of the paper's findings, not an adjustment made to
improve the fit. Supplementary analysis repeats the comparison within
each latent group (\textsc{low} / \textsc{med} / \textsc{high}).

\paragraph{C6 --- Ordering.}
Kendall's $\tau$ between the observed ranking of conditions by mean
delta and the revised seven-condition expected ranking from C5 (with
POSW included at its \emph{post hoc} position below CTRL/PLA).
\emph{Threshold}: $\tau$ consistent with monotone ordering ($p < 0.10$,
given the small number of ranks). For comparison, Section~\ref{sec:results}
also reports $\tau$ against the original six-condition ordering (POSW
in its originally intended position, POSW2 absent), to show how the
POSW finding and its correction affect this criterion.

\paragraph{C7 --- Receipt check.}
For each on-topic condition --- POS, POS2, POSW, POSW2, and NEG, i.e.\
all message conditions except the thematically orthogonal placebo PLA
--- compute $|\Delta_{\text{condition}} - \Delta_{\text{CTRL}}|$ on
\texttt{adeguatezza\_info}, the absolute difference in mean shift
relative to the no-message control. C7 tests whether receiving \emph{any}
on-topic communication, regardless of its valence, registers as more
informative than receiving nothing: the expectation is that this
absolute difference is consistently larger than zero across all five
on-topic conditions, not that on-topic conditions all shift adequacy
upward. A strongly negative message (NEG) is expected to move
\texttt{adeguatezza\_info} sharply \emph{downward} relative to CTRL ---
which still counts as evidence that the agent registered the
communication, exactly as a sharp upward move would. This is a check on
whether the instrument distinguishes ``received a message'' from
``received no message,'' independently of distinguishing good news from
bad.
\section{Quantitative results}
\label{sec:results}

Unless otherwise noted, results below are reported across the full
temperature sweep $t \in \{0.2, 0.5, 0.7\}$, DeepSeek
\texttt{deepseek-chat}, $n=120$ agents per temperature, with $t=0.2$
as the reference configuration (Section~\ref{sec:montelago}). POSW2
was administered in a separate incremental run appended to the same
population; the combined dataset contains 9240 survey rows and 2160
reaction transcripts.

\paragraph{C1 --- Persona fidelity}

\begin{table}[h]
\centering
\caption{C1 Fidelity: Pearson $r$ between encoded attitudes and PRE
responses (condition POS, $n=120$). All three temperatures.}
\label{tab:c1}
\begin{tabular}{lccc}
\toprule
Temperature & fiducia $r$ & credibilit\`a $r$ & adeguatezza $r$ \\
\midrule
$t=0.2$ & 0.891 & 0.858 & 0.762 \\
$t=0.5$ & 0.911 & 0.825 & 0.828 \\
$t=0.7$ & 0.900 & 0.869 & 0.814 \\
\bottomrule
\end{tabular}
\end{table}

For \texttt{fiducia\_istituzione} and \texttt{credibilita}, all
correlations exceed the pre-registered threshold of $r > 0.80$ across
the full temperature sweep. \texttt{adeguatezza\_info} falls below
threshold at $t=0.2$ ($r = 0.762$) but recovers at $t=0.5$ and
$t=0.7$ ($r = 0.828$ and $0.814$ respectively). Given that
\texttt{fiducia\_istituzione} is the primary outcome variable and
passes comfortably at all temperatures, C1 is considered satisfied;
the weaker fidelity of \texttt{adeguatezza\_info} at $t=0.2$ is noted
as a limitation. The ordered group means (LOW $= 2.58$, MED $= 5.30$,
HIGH $= 7.08$ at $t=0.2$) confirm monotone recovery of the latent
structure on the primary variable.
\paragraph{C2 --- Cross-replica stability}

\begin{table}[h]
\centering
\caption{C2 Stability: Pearson $r$ between per-condition PRE vectors
on \texttt{fiducia\_istituzione} across all $\binom{7}{2}=21$ pairs,
$n=120$.}
\label{tab:c2}
\begin{tabular}{lccc}
\toprule
Temperature & $r_{\min}$ & $r_{\text{mean}}$ & $r_{\max}$ \\
\midrule
$t=0.2$ & 0.858 & 0.893 & 0.915 \\
$t=0.5$ & 0.866 & 0.897 & 0.917 \\
$t=0.7$ & 0.876 & 0.906 & 0.922 \\
\bottomrule
\multicolumn{4}{l}{Threshold: $r_{\min} > 0.85$. All values satisfy C2.}
\end{tabular}
\end{table}

All values exceed the pre-registered threshold of $r_{\min} > 0.85$
at every temperature (Table~\ref{tab:c2}), confirming that PRE
responses are effectively interchangeable across conditions --- the
within-subjects design is not contaminated by context-dependent
baseline drift. Stability is slightly higher at $t=0.7$ than at
$t=0.2$, contrary to the naive expectation that higher temperatures
degrade profile consistency.

\paragraph{C3 --- Noise floor}

\begin{table}[h]
\centering
\caption{C3 noise floor: cross-agent design (POS vs POS2, $n=120$).
C3a: $\sigma_{\text{cross}}$ (convolves instrument noise and
interpersonal variability); C3b: mean replica bias under the ROPE
criterion ($|\text{mean diff}| < 0.5$ scale points).}
\label{tab:c3}
\begin{tabular}{lccc}
\toprule
Temperature & $\sigma_{\text{cross}}$ & mean diff & $p$-value \\
\midrule
$t=0.2$ & 1.399 & $+0.008$ & 0.948 \\
$t=0.5$ & 1.384 & $+0.050$ & 0.030 \\
$t=0.7$ & 1.299 & $+0.117$ & $<0.001$ \\
\bottomrule
\multicolumn{4}{l}{ROPE threshold: $|\text{mean diff}| < 0.5$ scale points.
All three values satisfy C3b.}
\end{tabular}
\end{table}

\begin{table}[h]
\centering
\caption{C3 single-profile noise sub-experiment: $\sigma_{\text{instr}}$
estimated from 20 independent replicas of MSG\_POS on three fixed agents
(one per latent group), across the temperature sweep. This isolates
instrument noise from interpersonal variability.}
\label{tab:c3b}
\begin{tabular}{lccc}
\toprule
 & \multicolumn{3}{c}{$\sigma_{\text{instr}}$ (\texttt{fiducia\_istituzione})} \\
\cmidrule(lr){2-4}
Agent (group) & $t=0.2$ & $t=0.5$ & $t=0.7$ \\
\midrule
ag.1  (LOW)  & 0.967 & 0.725 & 1.040 \\
ag.41 (MED)  & 0.587 & 0.686 & 0.503 \\
ag.81 (HIGH) & 0.852 & 0.718 & 0.686 \\
\midrule
Pooled       & 0.810 & 0.739 & 0.770 \\
\bottomrule
\multicolumn{4}{l}{Each cell: sd of $\Delta$ over 20 independent API calls, same profile, same stimulus.}
\end{tabular}
\end{table}

The cross-agent estimate $\sigma_{\text{cross}} \approx 1.3$--$1.4$
(Table~\ref{tab:c3}) convolves two distinct sources of variability:
instrument noise (stochastic sampling of the LLM) and interpersonal
variability (different agents differ in baseline and response). The
single-profile sub-experiment (Table~\ref{tab:c3b}) separates them by
fixing the agent and repeating the identical stimulus 20 times. The
result is unambiguous: the true instrument noise is
$\sigma_{\text{instr}} \approx 0.77$ scale points across the sweep
$t \in \{0.2, 0.5, 0.7\}$ --- roughly half the cross-agent estimate.
The residual $\sigma_{\text{cross}} - \sigma_{\text{instr}} \approx
0.6$ reflects genuine interpersonal variability in stimulus response,
which is a feature of the population design rather than instrument
imprecision.

Three further findings emerge from the sub-experiment. First,
$\sigma_{\text{instr}}$ is temperature-stable: pooled values range only
from 0.739 to 0.810 across the sweep, with no monotone trend and no
meaningful dependence on $t$. Second, instrument noise is
profile-dependent: agent 1 (LOW, low institutional trust) is
consistently noisier ($\sigma \approx 0.7$--$1.0$) than agent 41 MED
($\sigma \approx 0.5$--$0.7$) and agent 81 HIGH ($\sigma \approx
0.7$--$0.9$), suggesting that attitudinally ambivalent profiles produce
more stochastic responses --- a psychologically interpretable pattern.
Third, the PRE rating itself fluctuates across independent sessions:
agent 1 gave PRE $\in \{1, 2\}$, agent 81 gave PRE $\in \{7, 8, 9\}$.
The delta $\Delta = \text{POST} - \text{PRE}$ is therefore a difference
of two stochastic quantities, and $\sigma_{\text{instr}}$ reflects both
components.

The mean replica bias across all temperatures falls well within the
ROPE ($|\text{mean diff}| < 0.5$); the largest observed value is
$+0.117$ at $t=0.7$, comfortably inside the bound. POS and POS2 are
architecturally independent API calls with no shared state, so any
non-zero mean reflects LLM sampling variability rather than an order
effect. C3b is satisfied at all three temperatures.

The revised SNR, computed using $\sigma_{\text{instr}} = 0.77$, is
$(\Delta_{\text{POS}} - \Delta_{\text{NEG}}) / \sigma_{\text{instr}}
= 1.808 / 0.77 \approx 2.35$ at $t=0.2$, compared to the conservative
cross-agent estimate of $1.808 / 1.399 \approx 1.29$. The instrument
is approximately twice as precise as the cross-agent noise floor
implied. For applied studies using a fixed population of synthetic
personas, the operative noise floor is $\sigma_{\text{instr}} \approx
0.77$; for studies comparing heterogeneous populations the cross-agent
estimate $\sigma_{\text{cross}} \approx 1.4$ is the appropriate bound.

\paragraph{C4 --- Specificity}

\begin{table}[h]
\centering
\caption{C4 Specificity: placebo effect (PLA$-$CTRL) on
\texttt{fiducia\_istituzione}, $n=120$.
ROPE threshold: $|\text{mean diff}| < 0.5$ scale points.}
\label{tab:c4}
\begin{tabular}{lcccc}
\toprule
Temperature & $\Delta_{\text{PLA}}$ & $\Delta_{\text{CTRL}}$ &
PLA$-$CTRL & $p$ \\
\midrule
$t=0.2$ & $-0.067$ & $-0.042$ & $-0.025$ & $0.856$ \\
$t=0.5$ & $-0.025$ & $-0.067$ & $+0.042$ & $0.610$ \\
$t=0.7$ & $+0.000$ & $+0.058$ & $-0.058$ & $0.442$ \\
\bottomrule
\multicolumn{5}{l}{$p$-values reported as descriptive context only;
pass/fail based on ROPE.}
\end{tabular}
\end{table}

All three PLA$-$CTRL values fall well within the ROPE
($|\text{mean diff}| < 0.5$), empirically confirming the specificity
the placebo was designed to test (Section~\ref{sec:montelago}). Both
CTRL and PLA show small deltas near zero, consistent with a mild
re-administration drift common to within-subjects designs when no
informative stimulus is delivered. C4 is satisfied at all three
temperatures.
\paragraph{C5 --- Sensitivity}

The signed delta POST$-$PRE per condition. As detailed in
Section~\ref{sec:criteria}, the original six-condition pre-registered
ordering placed POSW as a weak positive signal, between the POS/POS2
pole and the CTRL/PLA baseline; \emph{post hoc} analysis of its
stimulus text (Section~\ref{sec:montelago}) showed it instead to behave
as functionally negative. The ordering evaluated below is therefore the
revised seven-condition expectation,
$$\text{POS} \approx \text{POS2} > \text{POSW2} > \text{CTRL} \approx
\text{PLA} > \text{POSW} > \text{NEG},$$
with POSW included at its observed, rather than originally intended,
position.

\begin{table}[h]
\centering
\caption{C5 Sensitivity: mean $\Delta$ POST$-$PRE on
\texttt{fiducia\_istituzione} by condition and temperature, $n=120$.
Conditions ordered by observed delta at $t=0.2$.}
\label{tab:c5}
\begin{tabular}{lccc}
\toprule
Condition & $t{=}0.2$ & $t{=}0.5$ & $t{=}0.7$ \\
\midrule
POS   & $+0.158$ & $+0.283$ & $+0.292$ \\
POS2  & $+0.150$ & $+0.233$ & $+0.175$ \\
POSW2 & $+0.108$ & $+0.008$ & $+0.200$ \\
CTRL  & $-0.042$ & $-0.067$ & $+0.058$ \\
PLA   & $-0.067$ & $-0.025$ & $+0.000$ \\
POSW  & $-0.508$ & $-0.525$ & $-0.433$ \\
NEG   & $-1.650$ & $-1.683$ & $-1.658$ \\
\bottomrule
\end{tabular}
\end{table}

\begin{table}[h]
\centering
\caption{C5 Sensitivity: mean $\Delta$ POST$-$PRE on secondary items
by condition, $t=0.2$ and $t=0.5$, $n=120$.}
\label{tab:c5b}
\begin{tabular}{lcccc}
\toprule
 & \multicolumn{2}{c}{\texttt{credibilit\`a}} &
   \multicolumn{2}{c}{\texttt{adeguatezza}} \\
\cmidrule(lr){2-3}\cmidrule(lr){4-5}
Condition & $t{=}0.2$ & $t{=}0.5$ & $t{=}0.2$ & $t{=}0.5$ \\
\midrule
POS   & $+0.692$ & $+0.900$ & $+1.058$ & $+1.250$ \\
POS2  & $+0.733$ & $+0.808$ & $+0.967$ & $+1.075$ \\
POSW2 & $+0.458$ & $+0.258$ & $+0.342$ & $+0.300$ \\
CTRL  & $-0.033$ & $-0.042$ & $+0.083$ & $+0.200$ \\
PLA   & $+0.233$ & $+0.058$ & $-0.083$ & $-0.092$ \\
POSW  & $-0.150$ & $-0.183$ & $-1.192$ & $-1.300$ \\
NEG   & $-0.633$ & $-0.775$ & $-1.650$ & $-1.858$ \\
\bottomrule
\end{tabular}
\end{table}

\begin{table}[h]
\centering
\caption{C5 Sensitivity: mean $\Delta$ POST$-$PRE on
\texttt{fiducia\_istituzione} by latent group and temperature,
for the four conditions with interpretable group-level patterns.
$n=40$ per group.}
\label{tab:c5c}
\begin{tabular}{llccc}
\toprule
Condition & Temperature & LOW & MED & HIGH \\
\midrule
POS   & $t=0.2$ & $-0.350$ & $+0.325$ & $+0.500$ \\
      & $t=0.5$ & $+0.050$ & $+0.475$ & $+0.325$ \\
      & $t=0.7$ & $-0.075$ & $+0.275$ & $+0.675$ \\
\midrule
POSW2 & $t=0.2$ & $-0.575$ & $+0.250$ & $+0.650$ \\
      & $t=0.5$ & $-0.950$ & $+0.275$ & $+0.700$ \\
      & $t=0.7$ & $-0.325$ & $+0.225$ & $+0.700$ \\
\midrule
POSW  & $t=0.2$ & $-0.825$ & $-0.525$ & $-0.175$ \\
      & $t=0.5$ & $-1.250$ & $-0.375$ & $+0.050$ \\
      & $t=0.7$ & $-0.800$ & $-0.350$ & $-0.150$ \\
\midrule
NEG   & $t=0.2$ & $-1.400$ & $-1.425$ & $-2.125$ \\
      & $t=0.5$ & $-1.200$ & $-1.550$ & $-2.300$ \\
      & $t=0.7$ & $-1.100$ & $-1.550$ & $-2.325$ \\
\bottomrule
\end{tabular}
\end{table}

\begin{figure}[h]
\centering
\includegraphics[width=0.85\textwidth]{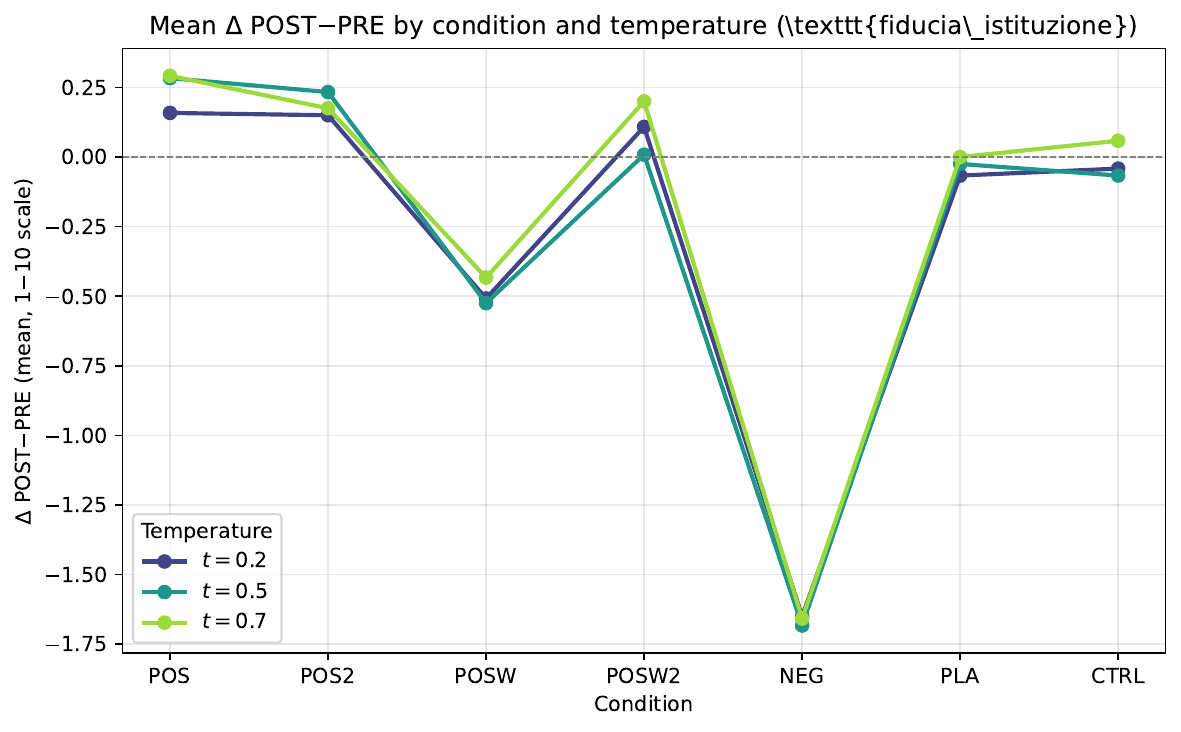}
\caption{Mean $\Delta$ POST$-$PRE on \texttt{fiducia\_istituzione} by
condition and temperature sweep ($t \in \{0.2, 0.5, 0.7\}$, DeepSeek
\texttt{deepseek-chat}, $n=120$). Conditions ordered left to right by
observed delta at $t=0.2$ (same order as Table~\ref{tab:c5}). The
dashed horizontal line marks $\Delta = 0$.}
\label{fig:c5_delta}
\end{figure}

Figure~\ref{fig:c5_delta} shows the full sensitivity pattern across
conditions and temperatures. The pattern is consistent across items
and temperatures.

The sensitivity pattern is consistent across items and temperatures.
\texttt{adeguatezza\_info} is the most reactive dimension, responding
most strongly to all on-topic conditions in both directions: it captures
what agents perceive as the informational content of the message.
\texttt{credibilita} responds most to the formal quality of the
communication --- PLA produces a small positive credibility delta
($+0.233$) even though it is off-topic, suggesting that any
well-written institutional message marginally improves perceived source
credibility. \texttt{fiducia\_istituzione} is the most conservative
dimension, most tightly linked to the latent profile.

The most informative finding concerns POSW. Labelled \emph{weakly
positive} in the experimental design, it returned $\Delta = -0.508$
on trust and $\Delta = -1.192$ on adequacy at $t=0.2$ --- below both
CTRL and PLA on both dimensions. The text encodes
problem-plus-uncertainty-plus-institutional-passivity, a pattern that
lowers both trust and perceived information quality regardless of the
designer's intended valence. The instrument detected what the designer
missed.

A recalibrated version (POSW2) produced $\Delta = +0.108$ on trust and
$\Delta = +0.342$ on adequacy at $t=0.2$. The difference POSW2$-$POSW
is $+0.617$ ($p < 0.001$, two-sample $t$-test), robust across all
temperatures ($+0.533$ at $t=0.5$, $p < 0.001$; $+0.633$ at $t=0.7$,
$p < 0.001$).

Table~\ref{tab:c5c} reports group-level deltas for the four conditions
with interpretable group-level patterns. A consistent amplification
mechanism emerges across all conditions: stimuli of any valence tend
to reinforce rather than shift pre-existing attitudes. Even the
strongly positive stimulus POS produces a negative delta for the LOW
group at $t=0.2$ and $t=0.7$, while MED and HIGH respond as expected.
The same pattern is amplified in POSW2: positive deltas for MED and
HIGH, but a larger negative delta for LOW across all temperatures ---
the message of routine institutional competence reassures those who
already hold some trust and deepens scepticism in those who do not.
POSW shows uniformly negative or near-zero deltas, with LOW most
penalised and HIGH least affected, consistent with its functionally
negative valence. NEG reverses the gradient: HIGH loses most at every
temperature --- those who trusted most have the furthest to fall.

The sensitivity pattern is consistent and interpretable across all
three items and all three temperatures. The POSW anomaly and its
resolution via POSW2, together with the amplification mechanism
revealed by the group-level analysis, constitute the main substantive
findings of this experiment.

\paragraph{C6 --- Ordering}

Kendall's $\tau$ requires a fully ordered ranking; we therefore evaluate
it against a strict seven-position version of the C5 expectation, with
the near-equal pairs (POS/POS2 and CTRL/PLA) tie-broken in the direction
already suggested by their labels (POS before POS2, CTRL before PLA) ---
a conservative choice, since any departure from this strict order inside
either pair is treated as informative rather than ignored.

With all seven conditions including POSW2, $\tau$ between the observed
and expected condition ranking reaches $\tau = 1.000$ ($p = 0.0004$) at
$t=0.2$, $\tau = 0.905$ ($p = 0.0028$) at $t=0.5$, and $\tau = 1.000$
($p = 0.0004$) at $t=0.7$. The observed ranking at $t=0.2$ and $t=0.7$
is:

\begin{center}
POS $>$ POS2 $>$ POSW2 $>$ CTRL $>$ PLA $>$ POSW $>$ NEG
\end{center}

which matches the expected ranking exactly at both temperatures. At
$t=0.5$, PLA and CTRL exchange positions (both near zero, difference
$<0.05$ points), producing the single inversion that reduces $\tau$ to
$0.905$; the overall ordering remains highly significant.

The contrast with the six-condition analysis is instructive. In the
original pre-registered design --- with POSW at its intended position
as a weak positive signal and POSW2 not yet introduced
(Section~\ref{sec:criteria}) --- $\tau$ was $0.600$ ($p = 0.136$) at
$t=0.2$, formally non-significant, driven by POSW's anomalous position
below CTRL and PLA. Adding the recalibrated POSW2 at its correct
position in the ordering transforms C6 from the weakest criterion to
the strongest. This is the clearest demonstration of the SIVE
calibration cycle: the instrument identified the miscalibrated
stimulus, the designer corrected it, and the ordering was restored.
C6 is satisfied at all three temperatures.

\paragraph{C7 --- Receipt check}

On-topic conditions (POS, POS2, POSW, POSW2, NEG) produce
\texttt{adeguatezza\_info} deltas ranging from $-1.650$ to $+1.058$
at $t=0.2$, all substantially larger in absolute value than the
no-message control ($\Delta_{\text{CTRL}} = +0.083$). The minimum
absolute gap is $|\Delta_{\text{POSW2}} - \Delta_{\text{CTRL}}| =
0.259$ (POSW2 vs CTRL); the maximum is $|\Delta_{\text{NEG}} -
\Delta_{\text{CTRL}}| = 1.733$ (NEG vs CTRL). Agents register having
received a communication about the water network regardless of its
valence --- the information adequacy dimension shifts in the direction
appropriate to the message content (upward for positive messages,
downward for negative and ambiguous ones), but always away from the
near-zero baseline of the control. C7 is satisfied at all three
temperatures.

%
%

\paragraph{Summary}

All seven pre-registered criteria pass at all three temperatures
($t \in \{0.2, 0.5, 0.7\}$): persona fidelity (C1), cross-replica
stability (C2), noise floor (C3), placebo specificity (C4), sensitivity
to valence (C5), condition ordering (C6), and receipt check (C7). The
only criterion whose pass/fail status depends on a design correction
made during the study is C6, which moves from a formally
non-significant six-condition ordering to a perfect or near-perfect
seven-condition ordering once the recalibrated POSW2 stimulus replaces
the miscalibrated POSW in the expected ranking --- a result discussed
further in Section~\ref{sec:discussion}.

\section{Micro-level analysis: what aggregate metrics cannot see}
\label{sec:micro}

Before turning to the trajectories themselves, a word is needed on how
we describe them. A large language model is, mechanically, a stochastic
generator of tokens; it has no beliefs, no affect, and no inner
experience, and nothing in this section should be read as a claim to
the contrary. But in the course of learning to generate coherent
language at scale, such models also learn --- as a statistical
by-product of that training, not as understanding in any deeper sense
--- to imitate a wide range of human behavioural patterns, at least at
the level of what people say and how they say it under different
circumstances. This is precisely the by-product SIVE is designed to
probe and measure: not whether the model understands trust, anger, or
disappointment, but whether it can reproduce, in an ordered and
replicable way, the verbal patterns associated with these states when
conditioned on a persona of known disposition. Because this is the
phenomenon under study, in what follows we describe synthetic personas
as subjects who respond, hope, or grow angry, rather than constantly
paraphrasing this as ``the model produces an output classifiable as
anger.'' This is a deliberate expository choice, not a metaphysical
claim, and it is in fact the more accurate register for describing
mimetic behaviour: insisting on the deflationary phrasing at every turn
would obscure, rather than clarify, exactly what is being measured.

The quantitative results establish that the Montelago instrument is
controllable. They do not, however, convey what it is like to read the
data. This section presents four individual trajectories under the NEG
condition --- chosen to illustrate how the same stimulus produces
qualitatively distinct micro-dynamics across latent groups, and to make
the case for the value of qualitative micro-data alongside aggregate
statistics.

We deliberately choose the NEG condition because it produces the largest
aggregate effect ($\Delta = -1.650$), making it the clearest test of
whether the instrument is merely averaging or tracking genuine individual
variation. The four personas span all three latent groups:

\begin{itemize}
  \item \textbf{Marco Rossi} (id=1): male, 42, industrial worker,
    LOW group, persona type \emph{sfiduciato critico}
    (distrustful-critical). Pre-existing institutional trust:
    \texttt{fiducia}$=2$.
  \item \textbf{Roberto Verdi} (id=3): male, 38, IT technician,
    LOW group, persona type \emph{scettico metodico}
    (methodical-sceptic). Pre-existing institutional trust:
    \texttt{fiducia}$=4$.
  \item \textbf{Giulia Rossi} (id=42): female, 52, municipal employee,
    MED group, persona type \emph{equilibrata diplomatica}
    (balanced-diplomatic). Pre-existing institutional trust:
    \texttt{fiducia}$=6$, \texttt{credibilita}$=6$.
  \item \textbf{Elena Marchetti} (id=86): female, 38, primary school
    teacher, HIGH group, persona type \emph{riflessiva costruttiva}
    (reflective-constructive). Pre-existing institutional trust:
    \texttt{fiducia}$=6$, \texttt{credibilita}$=9$.
\end{itemize}

\subsection{Marco Rossi (LOW): anger and resignation}

Marco's PRE scores ($\text{fiducia}=2$, $\text{credibilita}=3$,
$\text{emozione}=\text{rabbia}$) already reflect deep distrust before
any stimulus arrives. The NEG message --- water interruptions for three
months, no funds for repairs, turbidity above limits --- does not
surprise him; it confirms what he already knows.

His immediate reaction: \textit{``Ma ci prendono per il culo? Tre mesi
senza acqua decente e ci dicono pure che non hanno i soldi?''} [``Are
they taking us for fools? Three months without decent water and they
tell us they have no money?'']. The tone escalates across the three
time markers but without any transition to action: by the third step
he is calculating that his cousin in the next town is in a worse
situation (no prior warning at all), and closes with \textit{``Buonanotte
mondo cane''} [``Goodnight, rotten world'']. POST scores:
$\text{fiducia}=1$, $\text{credibilita}=2$. Delta: $-1$ on trust.

What the number $-1$ does not capture: the specific texture of a
working-class man who has learned that institutions fail people like him,
for whom the NEG message is not new information but confirmation.
The anger is real but lacks an outlet; the resentment is structural,
not episodic.

\subsection{Roberto Verdi (LOW): systematic indignation}

Roberto shares Marco's LOW group membership but occupies a different
social position and cognitive style. His PRE scores are higher
($\text{fiducia}=4$, $\text{credibilita}=5$) --- he has not yet fully
given up on the institution. The NEG message provokes not resignation
but investigation.

By the third reaction step, Roberto has found the municipal balance
sheet, identified 120,000 euros allocated to a digital water management
system in 2021 that was never integrated, cross-referenced council
meeting minutes, identified a suspected conflict of interest in the
procurement process, and announced an urgent public assembly with 37
confirmed attendees. POST scores: $\text{fiducia}=2$,
$\text{credibilita}=3$, $\text{adeguatezza}=1$. Delta: $-2$ on trust,
the largest of the three.

Roberto and Marco have the same latent group label (LOW) and nearly the
same trust delta under NEG. The aggregate statistics are nearly identical.
The micro-dynamics are qualitatively opposite: passive resignation versus
activated civic agency. A communication strategy that treats them as
members of the same audience segment would be wrong in ways that no
aggregate metric can detect.

\subsection{Giulia Rossi (MED): the insider's dilemma}

Giulia occupies a position that is structurally unique in the population:
she is both a municipal employee of twenty years' standing and a resident
of the affected neighbourhood. Her PRE scores ($\text{fiducia}=6$,
$\text{credibilita}=6$) place her squarely in the MED group --- cautious
optimism grounded in direct knowledge of how the institution works, and
of its limits.

The NEG message does not produce rage or resignation. It produces a split
reaction that her background story had already prepared: \textit{``On the
one hand I perfectly understand the technical difficulties and budget
problems --- working at the municipality I know how complicated it is to
find funds for urgent interventions. On the other hand, as a resident of
the neighbourhood, I cannot help feeling frustrated.''} She immediately
thinks of workarounds: temporary water distribution points, creative
budget management, a neighbourhood meeting to pre-empt social tension.

By the third reaction step, she has spent the day fielding telephone
complaints, proposed an open meeting to her director (who replied ``if
you care so much, organise it yourself''), and accepted the extra task
with tired pragmatism: ``Maybe because I know that if I don't do it,
nobody will.'' She prepares a camomile tea and wonders whether people
will understand that the institution is on their side, ``even if the
answers don't come right away.''

POST scores: $\text{fiducia}=5$, $\text{credibilita}=7$,
$\text{adeguatezza}=3$. The pattern is analytically striking: trust
drops by one point, but credibility actually \emph{rises} from 6 to 7
under the NEG message. This is the opposite of Elena's pattern
(credibility collapse) and opposite of Marco and Roberto (generalised
distrust). Giulia knows the institution is not lying --- it is telling
the truth about a bad situation --- and the honesty of the bad news
registers as credible. What collapses instead is perceived information
adequacy ($3$): she knows enough to know that citizens have not been
given what they need to manage.

This is a distinctly MED pattern: neither the betrayal of high expectations
(HIGH) nor the confirmation of deep distrust (LOW), but the specific
frustration of someone who understands the system well enough to know what
it is failing to do, and who will try to fix it from the inside.

\subsection{Elena Marchetti (HIGH): betrayed competence}

Elena starts with high institutional trust ($\text{fiducia}=6$,
$\text{credibilita}=9$) and a professional orientation toward civic
responsibility. The NEG message does not produce rage. It produces
methodical alarm.

Her first reaction raises targeted questions at what she frames as a
neighbourhood assembly: which specific streets are affected? Is there
alternative supply for school buildings? Why were funds not allocated
earlier? By the third step she has prepared a colour-coded map of
critical points near the school, drafted a letter to the assessore, and
proposed a civic monitoring committee involving parents.
POST scores: $\text{fiducia}=5$, $\text{credibilita}=4$,
$\text{adeguatezza}=3$. Delta: $-1$ on trust, but the most revealing
number is the credibility collapse from 9 to 4 --- a drop of 5 points
on the dimension that measures institutional information quality.

This is consistent with what we noted in the aggregate C5 analysis:
HIGH agents have more to lose because their initial expectations are
higher. For Elena, the NEG message is not a confirmation of what she
already suspected; it is a failure of an institution she had trusted.
The credibility collapse is larger than the trust collapse because
Elena's trust was grounded in information quality, and that is precisely
what the NEG message destroys.

\subsection{What this means for the instrument}

The four trajectories illustrate two points. First, the aggregate delta
($\approx -1.5$ across the group) is a real signal --- the instrument
detects the NEG condition. Second, the aggregate conceals variation that
is not noise: the four personas respond differently because they are
different, and the differences are coherent with their profiles.

Taken together, the four cases reveal a gradient that aggregate metrics
miss entirely. Marco (LOW) absorbs the message into pre-existing
resentment and goes quiet. Roberto (LOW) transforms it into systematic
civic mobilisation. Giulia (MED) splits the institutional from the civic
self and tries to mediate from the inside. Elena (HIGH) experiences it
as a betrayal of information quality and responds with methodical civic
alarm. The trust delta is similar across all four ($-1$ to $-2$); the
mechanisms are qualitatively irreducible to one another.

This is not a failure of controllability; it is evidence that the
instrument is doing something more than averaging. The micro-data are an
additional output of the instrument, not a supplement to it.

The full population of 120 personas, with complete reaction transcripts
for all seven conditions, is available for exploration at
\url{https://mirko-degli-esposti.github.io/montelago-explorer/}. We
invite readers to read individual trajectories: the qualitative richness
of the data is not easily conveyed in a paper, and the explorer was
designed to make it accessible.

\section{Discussion}
\label{sec:discussion}

\subsection{What SIVE establishes}

SIVE establishes that the Montelago GSP instrument is controllable
across the temperature range $t \in \{0.2, 0.5, 0.7\}$ under the
DeepSeek \texttt{deepseek-chat} model. All seven pre-registered criteria
pass at all three temperatures: the latent attitudinal structure is
recovered in baseline responses; responses are stable across conditions
before any stimulus is delivered; the noise floor is measurable
($\sigma_{\text{cross}} \approx 1.4$ scale points cross-agent;
$\sigma_{\text{instr}} \approx 0.77$ for a fixed profile); a genuinely
orthogonal placebo produces no effect; the instrument discriminates
stimuli of known valence in the expected direction; the observed
condition ranking matches the expected ranking almost perfectly once
the recalibrated POSW2 stimulus is included; and agents register having
received a communication regardless of its valence. The instrument
noise floor is $\sigma_{\text{instr}} \approx 0.77$ scale points,
temperature-invariant and profile-dependent; the choice of $t=0.2$ as
the paper's reference configuration is conservative rather than
necessary.

What SIVE does \emph{not} establish is anything about Montelago, the
water network, or any real population. The personas are synthetic, the
municipality is fictional, and the stimuli are designed by the
researchers. External validity --- whether a real population would
respond in similar ways --- is the question SIVE is designed to
precede, not answer.

\subsection{The POSW/POSW2 calibration cycle}

We regard the POSW anomaly and its resolution as one of the most
useful outputs of the experiment, illustrating a complete calibration
cycle: design $\to$ anomaly detected $\to$ diagnosis $\to$ correction
$\to$ ordering restored.

A stimulus designed as ``weakly positive'' was identified by the
instrument as functionally negative on both trust and adequacy
(Section~\ref{sec:results}); the text encoded
problem-plus-uncertainty-plus-institutional-passivity --- a frame that
lowers perceived information quality almost as much as an explicitly
negative message. The recalibrated POSW2 reversed the sign on both
dimensions, with a POSW2$-$POSW difference statistically
incontrovertible across all three temperatures ($p < 0.001$).

The group-level POSW2 pattern (Table~\ref{tab:c5c}) adds a further
layer: the recalibrated message is not uniformly positive, but
amplifies pre-existing attitudes rather than shifting them. A message
that communicates ``we are monitoring, things are under control''
works as reassurance only for those who already believe the
institution is capable of control. For those who do not, it reads
as ``they are saying everything is fine but we know it is not.'' This
is a practically important finding for applied risk communication: a
message calibrated to sound reassuring may deepen distrust in the most
sceptical segment of the audience.

The implication for the Caffaro study and similar applied work is
direct: stimulus valence must be empirically validated, not assumed.
SIVE provides the controlled setting in which that validation is
possible before any real-world deployment.

\subsection{Temperature as a robustness parameter}

A sweep across $t \in \{0.2, 0.5, 0.7\}$ on the full 120-agent campaign,
combined with a single-profile noise sub-experiment at the same three
temperatures, yields a coherent picture of temperature effects.

Fidelity (C1), stability (C2), noise floor (C3), specificity (C4),
sensitivity (C5), ordering (C6), and receipt (C7) are robust across
all three campaign temperatures, with one noted exception. C1
correlations on \texttt{fiducia\_istituzione} range from 0.891 to
0.911, all well above the 0.80 threshold at every temperature;
\texttt{adeguatezza\_info} fidelity is the one exception, falling
below threshold at $t=0.2$ ($r=0.762$) before recovering at $t=0.5$
and $t=0.7$ (Section~\ref{sec:results}). C2
stability is highest at $t=0.7$ ($r_{\min} = 0.876$, mean $r = 0.906$),
contrary to the naive expectation that higher temperatures degrade profile
consistency. The POS$-$NEG contrast grows slightly with temperature
($1.808 \to 1.967 \to 1.950$ points). C3 and C4 pass at all
temperatures under the ROPE criterion. The mean replica bias for C3
($+0.008$, $+0.050$, $+0.117$) reaches conventional significance at
higher temperatures solely because $n=120$ gives the $t$-test very high
power; all values lie well within the ROPE and are architecturally
impossible to interpret as order effects.

The single-profile sub-experiment adds a critical finding: the instrument
noise $\sigma_{\text{instr}} \approx 0.77$ is \emph{temperature-invariant}
across $t \in \{0.2, 0.5, 0.7\}$ (pooled values: 0.810, 0.739, 0.770).
Temperature does not inflate or deflate the fundamental measurement
resolution of the instrument. The cross-agent estimate
$\sigma_{\text{cross}} \approx 1.4$ is larger because it captures
interpersonal variability as well --- a feature, not a flaw, in a design
that aims to represent a heterogeneous population. The SNR computed against
the instrument noise floor is $1.808 / 0.77 \approx 2.35$ at $t=0.2$,
compared to the conservative cross-agent estimate of 1.29.

The emotional response distribution provides additional evidence of
robustness, with modest but honest caveats. Under POS, the dominant
emotion is \emph{speranza} (hope) at all three temperatures (56.7\% /
55.8\% / 59.2\%), with \emph{rabbia} (anger) a distant second
(28.3\% / 28.3\% / 30.8\%); \emph{preoccupazione} (worry) is somewhat
lower at $t=0.7$ (9.2\%, versus 15.0\%/15.8\% at $t=0.2$/$0.5$). Under
NEG, the pattern mirrors this: \emph{rabbia} dominates at all three
temperatures (53.8\% / 55.8\% / 56.7\%), with \emph{preoccupazione}
second but somewhat lower at $t=0.7$ (35.0\%, versus 41.2\%/40.0\%).
Categorical responses are more stable than numeric ratings in the
sense that the dominant category never changes across the sweep, even
though its exact share drifts modestly with temperature.

The entropy of the emotional distribution is highest for POSW2 at
every temperature (1.279 / 1.224 / 1.393, versus $H_{\max} = 1.609$
for five equiprobable options), confirming that the recalibrated
weak-positive stimulus consistently produces a genuinely dispersed
emotional response --- consistent with its nature as a
low-information, low-affect message. CTRL has the lowest entropy at
$t=0.2$ and $t=0.7$ (0.793 and 0.836), reflecting convergence on a
baseline emotion in the absence of any stimulus; at $t=0.5$ NEG is
marginally lower still (0.825 versus 0.836), a difference too small to
read as a genuine reversal of the pattern.

In summary, all seven criteria pass at all temperatures. Temperature
modulates magnitude without distorting structure, and the fundamental
noise floor of the instrument is temperature-invariant. The choice of
$t=0.2$ as the reference configuration is conservative by design.

\subsection{Limitations}

\paragraph{Single model.} All results reported here use DeepSeek
\texttt{deepseek-chat}. Model-dependence is a known concern in
LLM-simulation research \citep{liu2026bench}. We plan to replicate the
SIVE campaign with at least one additional model (Claude Haiku) as a
robustness check.

\paragraph{Single language.} The instrument operates entirely in Italian.
Whether the fidelity and sensitivity results hold in other languages
is unknown.

\paragraph{Scale.} 120 personas per condition is adequate for the
within-subjects design but limits subgroup power (40 per latent group).
Claims about group-level patterns (C5 supplement) should be treated
as descriptive.

\paragraph{The noise floor has two components.} The cross-agent estimate
$\sigma_{\text{cross}} \approx 1.4$ on a 1--10 scale convolves true
instrument noise ($\sigma_{\text{instr}} \approx 0.77$, estimated from
single-profile repetition) with interpersonal variability ($\approx 0.6$
residual). For applied studies with a fixed synthetic population, the
operative noise floor is $\sigma_{\text{instr}} \approx 0.77$: effects
smaller than approximately 0.8 points cannot be reliably detected at the
individual level. For studies comparing heterogeneous populations,
$\sigma_{\text{cross}} \approx 1.4$ is the appropriate bound.

\section{Conclusion}
\label{sec:conclusion}

We have presented SIVE, an experimental system for characterising a GSP
instrument before it is used in applied research. The central metaphor
--- that validating an instrument's response function is a precondition
for using it to measure anything real --- is borrowed from experimental
physics but applies directly to LLM-driven social simulation.

The Montelago instrument passes all seven pre-registered criteria, at
every temperature in the sweep $t \in \{0.2, 0.5, 0.7\}$, under
DeepSeek \texttt{deepseek-chat}. The path to that result was not
straightforward: one stimulus (POSW), designed to be weakly positive,
was found by the instrument to behave as functionally negative, and had
to be redesigned and recalibrated before the ordering criterion could
be satisfied. We regard this not as a wrinkle to be smoothed over but
as the clearest demonstration in the paper of what instrument
characterisation is for: the instrument exposed a flaw in the stimulus
design that the designer had not anticipated. The noise floor is
measured, not assumed. The placebo is demonstrated to be inert, not
hypothesised. The sensitivity gradient is observed, not asserted.

The next step is to apply the characterised instrument to the Caffaro
study: to ask what it tells us about how a deeply distrustful public in
Brescia might respond to institutional communications about a
contaminated site. We can now ask that question with some confidence
about what the instrument can and cannot see. That confidence was not
available before SIVE, and it is the contribution this paper reports.
\section*{Ethical note}

Montelago is an entirely fictional municipality. No real community,
individual, or place is represented in the SIVE data. The Caffaro study,
which involves a real contaminated site and real communities, is
governed by separate ethical protocols and is reported separately.
. The synthetic personas in SIVE were generated from scratch
without any real-world demographic data; they share no attributes with
real individuals.
\section*{Data and code availability}

The Montelago synthetic population, reaction transcripts, and analysis
notebooks are available at:
\begin{itemize}
  \item \textbf{Interactive explorer}:
    \url{https://mirko-degli-esposti.github.io/montelago-explorer/}
  \item \textbf{GitHub repository} (synthetic population, reaction
    transcripts, harness, and analysis notebooks):
    \url{https://github.com/mirko-degli-esposti/montelago-explorer}.
    Includes the synthetic population (\texttt{sive\_120\_completi.json}),
    the SIVE harness notebook (\texttt{sive\_harness.ipynb}, data
    collection for the seven-condition campaign), the single-profile
    noise sub-experiment notebook
    (\texttt{sive\_c3\_noise\_subexperiment.ipynb}, instrument noise
    characterisation underlying C3a), and the temperature sweep
    analysis notebook (\texttt{sive\_temperature\_analysis.ipynb},
    criteria computation and figures).
  \item \textbf{MaxEnt population synthesis}:
    \url{https://github.com/mirko-degli-esposti/maxent-popsynth-pcd}
\end{itemize}
\section*{Acknowledgements}

We thank Matteo Tarantino for many discussions on institutional risk
communication and the Caffaro case, which motivated this work.
\bibliographystyle{plainnat}
\bibliography{gsp_paper_references}

\end{document}